\documentclass[journal]{IEEEtran}
\usepackage[T1]{fontenc}
\usepackage{graphicx}
\usepackage{times}
\usepackage{helvet}
\usepackage{courier}
\usepackage{amsmath}
\usepackage{amssymb}
\usepackage{amsfonts}
\usepackage{multirow}
\usepackage{cite}
\usepackage{booktabs}
\usepackage{placeins}
\usepackage{float}
\usepackage{stfloats}

\usepackage{xcolor}

\definecolor{citecolor}{HTML}{0071BC}
\definecolor{SeaGreen4}{RGB}{0,205,102}
\definecolor{SlateBlue}{RGB}{106,90,205}
\definecolor{DarkRed}{RGB}{178,34,34}

\usepackage[
    colorlinks=true,
    linkcolor=black,
    citecolor=citecolor,
    urlcolor=citecolor
]{hyperref}

\usepackage{pifont}

\begin{document}

\title{Nuclear Fusion Knowledge Graph Guided Edge-Localized Mode Recognition on EAST}

\author{Wanli Lyu, Laiqi Qian, Xiao Wang$^{*}$, Qingquan Yang$^{*}$, Zehua Chen, Dengdi Sun, Guosheng Xu, Jin Tang\\
\thanks{$\bullet$ Wanli Lyu, Laiqi Qian, Xiao Wang and Jin Tang are with the School of Computer Science and Technology, Anhui University, Hefei 230601, China. (e-mail: \{xiaowang, lwl, tangjin\}@ahu.edu.cn, 1784291876@qq.com)}  
\thanks{$\bullet$ Dengdi Sun is with School of Computer Science and Technology, Anhui University, Hefei 230601, China (email: sundengdi@ahu.edu.cn)}
\thanks{$\bullet$ Zehua Chen is with the Department of Computer Science and Technology, Tsinghua University, Beijing, China. (email: zhc23thuml@tsinghua.edu.cn)}
\thanks{$\bullet$ Qingquan Yang and Guosheng Xu are with the Institute of Plasma Physics, Chinese Academy of Sciences, Hefei 230031, China (email: \{yangqq, gsxu\}@ipp.ac.cn)} 
\thanks{* Corresponding authors: Xiao Wang $\&$ Qingquan Yang}
}

\maketitle
\begin{abstract}
Edge-localized modes (ELMs) are transient magnetohydrodynamic instabilities commonly observed in tokamak plasmas. Their repetitive bursts impose intense transient heat and particle loads on plasma-facing components, making accurate recognition of different ELM-related plasma states important for plasma-state monitoring, safe operation, and active control of fusion devices. Although deep time-series models can learn complex dynamic patterns from multi-diagnostic signals, their predictions are mainly data-driven and lack explicit utilization of ELM physics, plasma operating states, and historical discharge experience. As a result, they may confuse categories with similar temporal patterns, particularly under severe class imbalance. To address this issue, we propose KGF-ELM, a Knowledge Graph Fusion Network for ELM Recognition, and introduce a Knowledge Graph Collaborative Fusion Module (KGCFM) that organizes physical evidence from the current time window, ELM prototypes, relational knowledge graph representations, and similar historical discharge cases into complementary multi-source knowledge representations. KGCFM collaboratively models these knowledge sources through physics prototype matching, graph-structured knowledge encoding, historical case retrieval, cross-evidence consistency analysis, and reliability estimation, thereby generating knowledge representations for enhancing temporal features and correcting classification results. Based on this module, KGF-ELM employs a dual-level reliability-aware fusion mechanism. At the feature level, physical residuals and graph-guided knowledge representations are adaptively incorporated into temporal features according to their reliability. At the decision level, reliability gating, minority-class protection, and bounded knowledge correction are used to safely refine the classification results and reduce the negative influence of unreliable knowledge on the base model. Experiments on EAST-ELM640 demonstrate that KGF-ELM consistently
improves five temporal backbones, yielding an average Macro-F1 gain of
4.74 percentage points, while the best configuration achieves 91.7\%
Accuracy, 81.6\% Macro-F1, and 98.0\% Macro-AUC.
\end{abstract}


\section{Introduction}
\label{sec:introduction}

\IEEEPARstart{M}{agnetic confinement} fusion is widely regarded as a promising pathway toward sustainable and low-carbon energy production. Tokamaks commonly operate in high-confinement mode (H-mode) to achieve improved plasma confinement; however, H-mode operation is often accompanied by edge-localized modes (ELMs)~\cite{zohm1996edge,leonard2014edge,snyder2002edge}. ELMs are transient edge instabilities that repetitively release particles and energy from the confined plasma, imposing substantial transient loads on plasma-facing components and posing challenges for future long-pulse fusion devices~\cite{loarte2003elm}. Correspondingly, ELM mitigation and active control have been extensively investigated using magnetic perturbations and related plasma-control strategies~\cite{evans2005suppression,liang2007active}. Experiments on the Experimental Advanced Superconducting Tokamak (EAST) have further revealed diverse edge behaviors and operating regimes, including small-ELM, edge-coherent-mode, quiescent high-confinement, and actively mitigated ELM conditions~\cite{wang2014edge,hu2015new,lin2025edge}; nonlinear simulations have also been used to analyze ELM dynamics on EAST~\cite{liu2014three}. With the continued development of multi-diagnostic systems and the accumulation of experimental data, accurately recognizing different ELM-related plasma states from multivariate time-series signals has become an important task for intelligent plasma diagnosis, operational-state monitoring, and the development of future ELM control strategies.

Recent deep temporal models have demonstrated strong capabilities for multivariate time-series analysis through patch-based representation learning, multi-periodic modeling, and variable-wise temporal encoding~\cite{nie2022time,wu2022timesnet,liu2024itransformer}. Such models can capture local burst patterns, long-range temporal dependencies, and cross-channel correlations from multiple diagnostic signals, providing a powerful data-driven foundation for plasma-state recognition. Nevertheless, their predictions are primarily determined by statistical patterns learned from training data, while ELM physics, plasma operating conditions, and historical discharge experience are generally not represented explicitly. This limitation is particularly relevant to fine-grained ELM recognition, because different ELM-related states may exhibit partially similar local waveforms despite corresponding to distinct physical conditions. Meanwhile, Type-III ELM and Large ELM states are substantially less represented than the dominant operating states, making it more difficult for purely data-driven models to learn balanced decision boundaries, a general difficulty also emphasized by class-balanced learning studies~\cite{cui2019class}. Variations among discharges and diagnostic uncertainty further increase the ambiguity of recognition under complex plasma conditions.

ELM recognition, however, is supported by multiple forms of domain knowledge that are complementary to temporal representations learned directly from data. Physical observations such as $D_{\alpha}$ bursts, edge-density responses, plasma stored-energy variations, confinement conditions, and cross-diagnostic consistency provide important evidence for distinguishing ELM-related states~\cite{leonard2014edge,wang2014edge,teo2024using}; dedicated optical diagnostics on EAST further provide direct measurements of characteristic plasma-emission signals~\cite{xu2016filterscope}. Recurring combinations of these physical characteristics can be summarized as ELM prototypes that describe class-level patterns, while representative historical discharge windows provide instance-level experience from previously observed plasma conditions. Knowledge graphs offer a natural means of organizing such heterogeneous information by connecting physical evidence, ELM prototypes, and class semantics through typed relations~\cite{ji2021survey,hogan2021knowledge}. Combining relational knowledge with historical case memory therefore provides an opportunity to complement temporal representations with both structured physical priors and experience derived from previous discharges.

Directly incorporating such knowledge into a temporal model is nevertheless nontrivial. Physical indicators can become uncertain under noisy or transitional conditions, prototype support may be ambiguous, and retrieved historical cases may only partially match the current plasma state. Different knowledge sources may consequently provide inconsistent or even conflicting evidence. Moreover, predictive confidence from modern neural networks is not necessarily well calibrated~\cite{guo2017calibration}, which makes an unconstrained confidence-based correction particularly risky. If these signals are injected into the temporal model without appropriate control, unreliable knowledge may override an already reliable data-driven prediction and can be particularly harmful to underrepresented ELM classes. In addition, historical knowledge construction and retrieval must be carefully isolated from validation and test discharges to avoid information leakage. Therefore, the central challenge is not simply how to introduce additional domain knowledge, but how to determine which knowledge is relevant and reliable for the current diagnostic window and how strongly it should influence the final prediction. This motivates a knowledge-enhanced framework in which the temporal model remains the primary prediction pathway, while structured knowledge is treated as reliability-weighted complementary evidence.

To address these challenges, we propose the \emph{Knowledge Graph Fusion Network for ELM Recognition} (KGF-ELM), a unified framework for point-wise recognition of four ELM-related plasma states on EAST. KGF-ELM consists of a replaceable temporal backbone and a \emph{Knowledge Graph Collaborative Fusion Module} (KGCFM). A discharge-isolated ELM knowledge resource organizes physical evidence, ELM prototypes, typed relational knowledge, and representative historical cases, with historical memory constructed exclusively from training discharges and training-time retrieval excluding cases from the query discharge. During inference, the temporal backbone first extracts data-driven representations and produces base predictions. In parallel, KGCFM evaluates physical evidence from the current diagnostic window, activates sample-relevant relational knowledge, retrieves similar historical cases, and coordinates these complementary sources through cross-source consistency analysis and reliability estimation. The resulting knowledge representations are then integrated through a reliability-aware dual-level fusion mechanism: reliable physics and graph--case information enhance temporal representations at the feature level, while a bounded knowledge correction refines uncertain predictions at the decision level. A minority-class protection mechanism further constrains potentially harmful corrections to Type-III ELM and Large ELM predictions. Experiments on EAST-ELM640 demonstrate consistent improvements in Macro-F1 and recall across multiple temporal backbones, indicating that KGCFM can provide complementary knowledge enhancement without being tied to a particular temporal architecture.

The main contributions of this work are summarized as follows:
\begin{itemize}
\item We propose KGF-ELM, a knowledge-enhanced framework that combines multi-diagnostic temporal modeling with structured physical knowledge for point-wise recognition of four ELM-related plasma states on EAST. Its replaceable temporal backbone enables the proposed knowledge pathway to be integrated with different time-series representations.

\item We construct a discharge-isolated ELM knowledge resource that jointly organizes physical evidence, ELM prototypes, a typed relational graph, and a training-only historical case memory. Discharge-level isolation prevents validation/test leakage, while
same-shot exclusion during training-time retrieval avoids trivial
matches from highly overlapping windows of the same discharge.
Relational and instance-level knowledge can therefore complement the
current diagnostic window under a training-only knowledge protocol.

\item We develop KGCFM together with a reliability-aware dual-level fusion mechanism that coordinates physical, graph, and historical knowledge through consistency and reliability estimation. The proposed framework enhances temporal representations at the feature level and performs bounded decision-level correction with minority-class protection, thereby reducing the adverse influence of unreliable or conflicting knowledge.

\end{itemize}

\section{Related Work}
\label{sec:related_work}

\subsection{Multivariate Time-Series Modeling}

Deep learning has become a dominant approach for multivariate time-series modeling. Convolutional neural networks capture local temporal patterns through shared convolutional filters, while recurrent neural networks model sequential dependencies through hidden-state transitions. Transformer-based architectures further improve long-range modeling by using self-attention to capture dependencies across temporal positions and variables~\cite{vaswani2017attention}. Early long-sequence forecasting models improve efficiency and temporal structure modeling through sparse attention, decomposition, autocorrelation, and frequency-domain representations~\cite{zhou2021informer,wu2021autoformer,zhou2022fedformer}. More recently, a variety of specialized time-series architectures have been developed to improve representation efficiency and capture different forms of temporal structure. PatchTST divides time series into local patches to construct compact temporal representations, TimesNet models complex temporal variations through multi-periodicity, and iTransformer performs variable-wise tokenization to explicitly capture multivariate correlations~\cite{nie2022time,wu2022timesnet,liu2024itransformer}. Lightweight approaches such as DLinear have also demonstrated that simple temporal decomposition and linear mapping can remain competitive for time-series modeling~\cite{zeng2023transformers}.

Beyond these architectures, recent models have explored increasingly diverse temporal representation mechanisms. Timer introduces generative pre-training for large-scale time-series modeling, while Timer-XL extends this direction toward long-context unified time-series analysis~\cite{liu2024timer,liu2025timer}. TimeMixer performs decomposable multiscale mixing, whereas SOFTS models multivariate interactions through series-core fusion~\cite{wang2024timemixer,han2024softs}. TimeXer explicitly incorporates exogenous-variable information, TimeFilter models patch-specific spatial--temporal dependencies through graph-based filtration, and CrossLinear captures cross-variable correlations through a lightweight embedding mechanism~\cite{wang2024timexer,hu2025timefilter,zhou2025crosslinear}. These models provide strong and complementary temporal encoders, but their predictions remain primarily driven by statistical relationships learned from numerical observations. The physical meanings of diagnostic signals, plasma operating conditions, and historical discharge experience are generally not explicitly represented during inference. Rather than designing another task-specific temporal architecture, this work investigates whether structured physical and historical knowledge can serve as a complementary source of information for different temporal backbones.

\subsection{Artificial Intelligence for Plasma Diagnosis and ELM Recognition}

The increasing availability of diagnostic data from magnetic-confinement fusion experiments has motivated extensive applications of machine learning and deep learning to plasma analysis. Data-driven approaches have been investigated for disruption prediction, unstable-event identification, and magnetic plasma control~\cite{kates2019predicting,rea2018disruption,churchill2020deep,fu2020machine,zhu2023integrated,degrave2022magnetic}, while automated plasma-state and ELM analysis has also become an active research direction. Compared with manually designed thresholds and handcrafted rules, learned models can extract complex nonlinear patterns directly from multi-diagnostic measurements and provide a flexible way to analyze high-dimensional plasma dynamics.

For ELM analysis, previous studies have investigated automatic ELM classification on JET~\cite{duro2009automated}, confinement-state and ELM recognition on TCV and KSTAR~\cite{matos2020classification,shin2020real}, and automatic ELM identification or detection on DIII-D~\cite{o2023automatic,teo2024using}. These studies primarily address ELM-event detection, ELM classification, or broader confinement-state recognition. Fine-grained point-wise recognition of different ELM-related states remains challenging because their diagnostic signatures may overlap and their class distributions can be highly imbalanced. Conventional physical analysis relies on characteristic indicators such as $D_{\alpha}$ emission, density responses, plasma stored energy, confinement conditions, and magnetic or radiative diagnostics to characterize ELM activity. In most data-driven pipelines, however, such information is either treated as ordinary numerical input or converted into manually designed criteria, rather than being organized as reusable structured knowledge and explicitly coordinated with learned temporal representations. Historical discharge experience is likewise rarely incorporated as an instance-level reference during prediction. These limitations motivate a framework that can retain the representation capability of modern temporal models while explicitly exploiting physical evidence, recurring ELM patterns, relational structure, and similar historical cases.

\subsection{Physics-Informed Learning and Structured Knowledge Reasoning}

Physics-informed learning aims to incorporate physical laws, constraints, or domain priors into data-driven models in order to improve their physical consistency and generalization~\cite{raissi2019physics,karniadakis2021physics}. More broadly, scientific knowledge can be integrated with machine learning through constraints, architecture design, hybrid modeling, or auxiliary domain representations~\cite{willard2022integrating}. Classical physics-informed neural networks typically introduce governing equations, boundary conditions, or other analytical constraints into network optimization. For complex diagnostic tasks in which a complete governing equation is difficult to incorporate directly, domain knowledge can instead be represented through observable physical evidence and structured semantic relationships. In ELM recognition, diagnostic responses extracted from the current plasma window provide sample-dependent physical evidence, while recurring combinations of physical characteristics can be represented as class-level prototypes. Prototype-based learning provides compact class references for comparison and reasoning~\cite{snell2017prototypical}.

Knowledge graphs provide another mechanism for organizing heterogeneous domain information through entities and typed relations~\cite{ji2021survey,hogan2021knowledge}. Relational graph convolutional networks enable information propagation over multi-relational graph structures~\cite{schlichtkrull2018modeling}, while attention-based graph learning provides a general mechanism for assigning different importance to neighboring graph elements~\cite{velivckovic2017graph}. In parallel, retrieval-augmented methods demonstrate the value of dynamically accessing external evidence instead of encoding all useful information in fixed model parameters~\cite{lewis2020retrieval,karpukhin2020dense,guu2020retrieval}. Recent retrieval-augmented time-series modeling further shows that historical temporal patterns can serve as useful contextual references~\cite{han2025retrieval}. These forms of knowledge operate at different semantic levels: physical evidence describes the current diagnostic window, prototypes represent recurring class-level patterns, graph relations encode structural dependencies, and retrieved cases provide historical operating experience. Treating them as a single homogeneous representation or directly concatenating their features does not account for differences in reliability, granularity, or sample relevance.

Moreover, the reliability of domain knowledge can vary across individual plasma windows. Diagnostic noise may weaken physical evidence, transition regions may activate multiple competing prototypes, and retrieved historical cases may be only partially consistent with the current operating state. Consequently, knowledge-enhanced prediction requires not only structured knowledge representation but also sample-specific coordination of complementary or conflicting knowledge sources. KGF-ELM addresses this problem by constructing a discharge-isolated ELM knowledge resource and introducing KGCFM to coordinate physical evidence, ELM prototypes, relational graph representations, and training-derived historical cases. Their influence on temporal representations and final decisions is regulated through cross-source consistency and reliability-aware fusion, enabling structured knowledge to complement rather than indiscriminately override the data-driven temporal pathway.

\section{Our Proposed Approach}
\label{sec:method}

\begin{figure*}[t]
    \centering
    \includegraphics[width=0.99\textwidth]
    {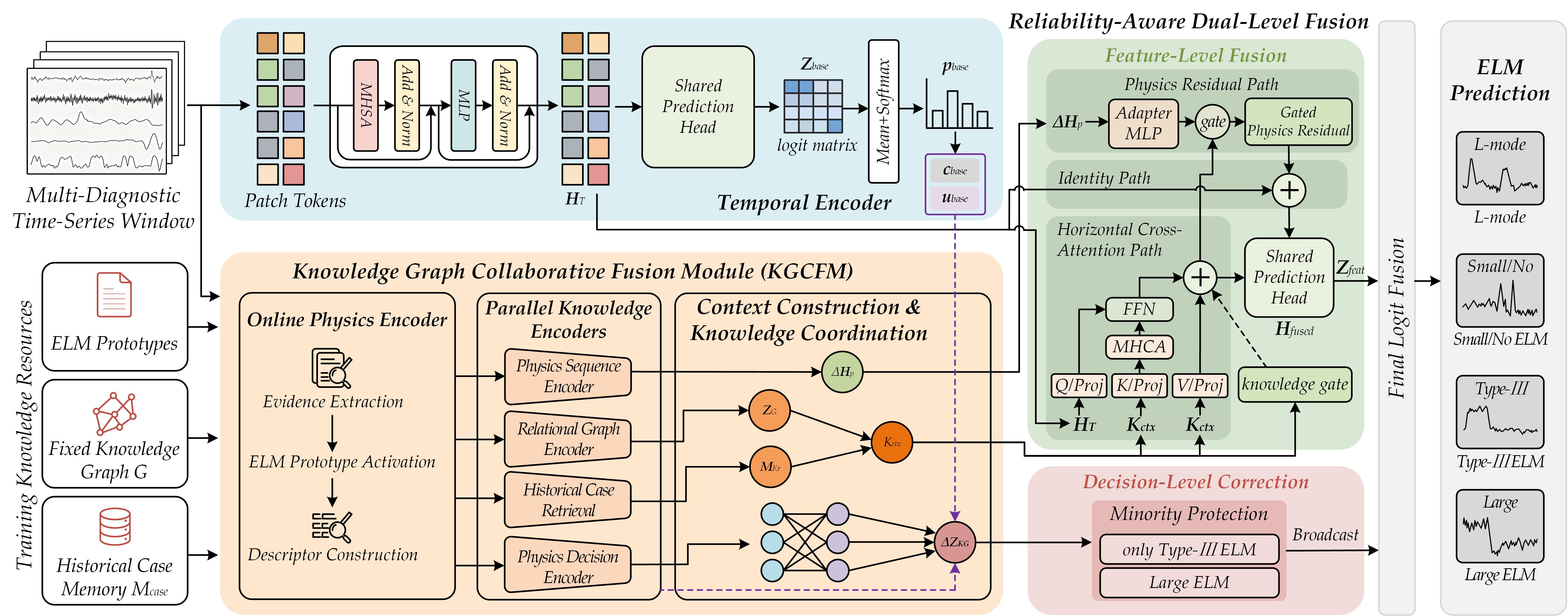}
    \caption{Overall architecture of KGF-ELM. The temporal backbone
    extracts data-driven temporal representations and produces
    point-wise base predictions. In parallel, KGCFM evaluates current
    physical evidence, pools relational graph knowledge, and retrieves
    similar training cases to generate a physics residual, a
    graph--case context, and a reliability-aware decision correction.
    These knowledge signals are incorporated through feature-level
    fusion and bounded decision-level correction.}
    \label{fig:overall_framework}
\end{figure*}

KGF-ELM is designed to combine data-driven temporal modeling with
explicit domain knowledge from fusion physics without forcing either source to
dominate the prediction. As illustrated in
Fig.~\ref{fig:overall_framework}, the framework consists of a
replaceable temporal backbone and a Knowledge Graph Collaborative
Fusion Module (KGCFM). The temporal branch learns diagnostic dynamics
directly from multichannel measurements and provides point-wise base
predictions. In parallel, KGCFM analyzes physical evidence in the
current window, activates a typed ELM knowledge graph, and retrieves
similar historical cases from training discharges. These complementary
signals are coordinated according to their reliability and then fused
with the temporal representation at both feature and decision levels.
This separation is intentional: the backbone remains responsible for
fine-grained temporal discrimination, whereas domain knowledge acts as
an adaptive source of evidence when purely data-driven predictions are
uncertain or physically ambiguous.

\subsection{Task Formulation}
\label{subsec:task_formulation}

Given a batch of multivariate diagnostic windows, the input is
\begin{equation}
    \boldsymbol{X}\in\mathbb{R}^{B\times T\times C},
\end{equation}
where $B$, $T$, and $C$ denote the batch size, window length, and
number of diagnostic channels, respectively. The four-class label set
$\mathcal{Y}=\{0,1,2,3\}$ corresponds to L-mode, Small/No ELM,
Type-III ELM, and Large ELM. Unlike window-level event recognition,
our objective is to preserve the temporal position of every prediction.
The point-wise targets and model outputs are therefore defined as
\begin{equation}
\begin{aligned}
    \boldsymbol{Y}&=[y_{b,t}]\in\mathcal{Y}^{B\times T},
    \qquad K=|\mathcal{Y}|=4,\\
    \boldsymbol{P}_{\mathrm{final}}
    &=F(\boldsymbol{X})\in[0,1]^{B\times T\times K},\\
    \widehat{y}_{b,t}
    &=\operatorname*{argmax}_{k\in\mathcal{Y}}
      P_{\mathrm{final},b,t,k}.
\end{aligned}
\label{eq:task_definition}
\end{equation}

This formulation is important for ELM analysis because a single
window may contain transitions or short-lived signatures whose timing
would be lost under a single window label. It also introduces two
difficulties. First, different ELM states can exhibit partially
overlapping local waveforms, so a temporally plausible prediction is
not necessarily consistent with fusion physics. Second, Type-III and
Large ELM samples are substantially less frequent than the dominant
states, making their decision boundaries more sensitive to noisy or
ambiguous evidence. KGF-ELM addresses these issues without changing the
point-wise task: the temporal branch learns the primary discriminative
representation, while physical rules, relational knowledge, and
historical cases provide complementary evidence whose influence is
explicitly reliability controlled.

\subsection{Temporal Encoding and Base Prediction}
\label{subsec:temporal_encoding}

We deliberately keep the temporal component backbone-agnostic so that
the benefit of knowledge guidance is not tied to a particular sequence
architecture. For an input window $\boldsymbol{X}$, a selected temporal
encoder $f_{\theta}$ produces
\begin{equation}
    \boldsymbol{H}_{T}=f_{\theta}(\boldsymbol{X}),
    \qquad
    \boldsymbol{H}_{T}\in\mathbb{R}^{B\times N\times d},
\label{eq:temporal_encoding}
\end{equation}
where $N$ is the number of temporal tokens and $d$ is the hidden
dimension. Because different backbones may preserve, downsample, or
patch the original sequence differently, $N$ is not required to equal
$T$. A shared prediction interface uses the hidden representation and
the backbone-specific decoding context $\mathcal{C}_{T}$ to recover
time-aligned logits,
\begin{equation}
    \boldsymbol{Z}_{\mathrm{base}}
    =h_{\psi}(\boldsymbol{H}_{T},\mathcal{C}_{T})
    \in\mathbb{R}^{B\times T\times K}.
\label{eq:base_logits}
\end{equation}
This interface lets KGCFM operate on a common token representation
while leaving the native temporal modeling mechanism of each backbone
unchanged.

Besides point-wise logits, knowledge coordination requires a compact
estimate of how decisive the data-driven branch is for the current
window. We obtain a window-level base distribution only for this
purpose:
\begin{equation}
    \boldsymbol{p}_{\mathrm{base}}
    =\operatorname{Softmax}\!\left(
    \frac{1}{T}\sum_{t=1}^{T}
    \boldsymbol{Z}_{\mathrm{base}}[:,t,:]\right).
\label{eq:base_distribution}
\end{equation}
The corresponding window-level base prediction is defined as
$\widehat{y}^{\,\mathrm{base}}_b
=\operatorname*{argmax}_{k}p_{\mathrm{base},b,k}$.
The maximum probability gives the base confidence
$c_{\mathrm{base}}$, and the normalized entropy yields an uncertainty
measure $u_{\mathrm{base}}$. These quantities are not auxiliary
classification targets and do not convert the task into window-level
recognition. Instead, they inform the later gates whether knowledge
correction is necessary. When the backbone is already confident,
knowledge injection can be conservative; when the base distribution is
uncertain, physically consistent evidence can receive more influence.
For EAST-ELM640, all backbones use $T=100$, $C=18$, and $K=4$.

\begin{figure*}[t]
    \centering
    \includegraphics[width=0.99\textwidth]
    {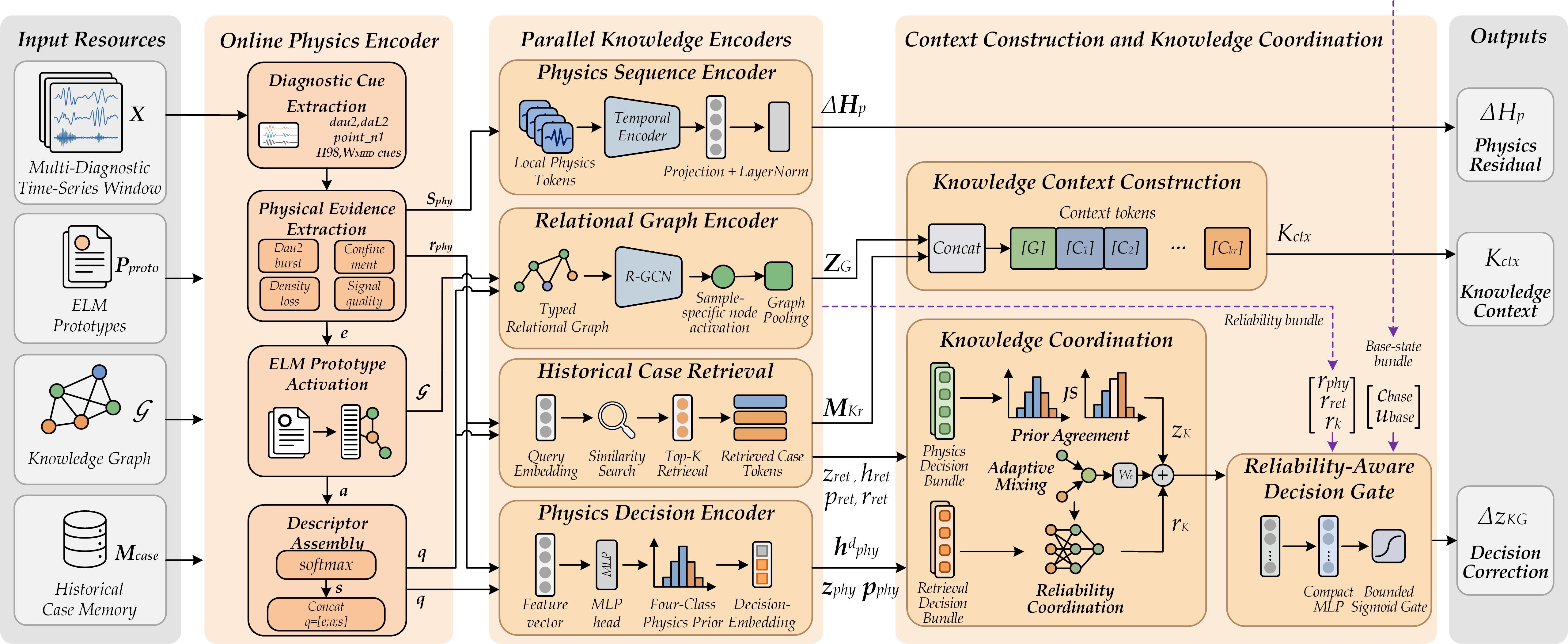}
    \caption{Detailed architecture of KGCFM. The online physics
    analyzer evaluates fixed evidence and prototype definitions from
    the current window. Physics, graph, and retrieval pathways produce
    a physics residual, relational graph token, historical-case
    representations, and reliability cues. Graph and case tokens form
    $\boldsymbol{K}_{\mathrm{ctx}}$, while coordinated knowledge and
    reliability gates generate the bounded decision correction
    $\Delta\boldsymbol{z}_{\mathrm{KG}}$.}
    \label{fig:kgcfm_framework}
\end{figure*}

\subsection{ELM Knowledge Resource Construction}
\label{subsec:knowledge_construction}

The knowledge resource is designed to represent three complementary
forms of information: interpretable physical evidence, stable
relations among ELM concepts, and empirical precedents from historical
discharges. We write
\begin{equation}
    \mathcal{K}_{\mathrm{ELM}}
    =\{\mathcal{E},\mathcal{P},\mathcal{G},
      \mathcal{M}_{\mathrm{case}},\mathcal{I}_{\mathrm{ret}}\},
\label{eq:knowledge_resource}
\end{equation}
where $\mathcal{E}$ and $\mathcal{P}$ contain physical-evidence and
ELM-prototype definitions, $\mathcal{G}$ is a typed relational graph,
$\mathcal{M}_{\mathrm{case}}$ stores representative historical cases,
and $\mathcal{I}_{\mathrm{ret}}$ is the corresponding retrieval index.
The physical-evidence definitions, deterministic prototype definitions,
graph topology, relation types, and semantic edge weights are fixed
after offline construction. The graph schema is validated using
structured records from the designated training-only construction
subset, while the historical case memory and retrieval index are
constructed exclusively from this subset.

\noindent\textbf{Physical evidence and prototype activation.}
Rather than feeding symbolic rules directly into the classifier, we
first convert the current diagnostic window into a continuous knowledge
descriptor. The online analyzer evaluates 11 physics-aware evidence
terms that summarize characteristic behavior of the available
diagnostics and compares them with 15 ELM-related prototypes. The
prototype layer provides a soft description of how closely the current
window resembles known physical patterns instead of making a hard
class assignment. Four class-support values then aggregate the
evidence into class-oriented physical support. The resulting descriptor
is
\begin{equation}
    \boldsymbol{q}
    =[\boldsymbol{e};\boldsymbol{a};\boldsymbol{s}]
    \in[0,1]^{30},
\label{eq:knowledge_descriptor}
\end{equation}
where $\boldsymbol{e}\in\mathbb{R}^{11}$,
$\boldsymbol{a}\in\mathbb{R}^{15}$, and
$\boldsymbol{s}\in\mathbb{R}^{4}$ denote evidence, prototype
activation, and class support, respectively. Importantly,
$\boldsymbol{s}$ is produced from physical evidence and is used only as
a soft prior; the ground-truth label of the query is never inserted
into $\boldsymbol{q}$. This avoids turning the knowledge pathway into a
label shortcut.

\noindent\textbf{Typed relational graph.}
The same 30 semantic quantities form the nodes of a compact
computational graph containing 11 evidence nodes, 15 prototype nodes,
and four class nodes. We connect them with 100 directed weighted edges
of four relation types: \texttt{supports}, \texttt{contradicts},
\texttt{indicates}, and \texttt{associated\_with}. Relation types
distinguish qualitatively different physical interactions, while edge
weights encode the relative strength of predefined semantic links.
The topology and semantic weights remain fixed during training; only
the neural parameters used to propagate and pool graph information are
optimized. Consequently, the model can learn how to use the graph
without rewriting the physical meaning of its relations.

\begin{figure*}[t]
\centering
\includegraphics[width=\textwidth]{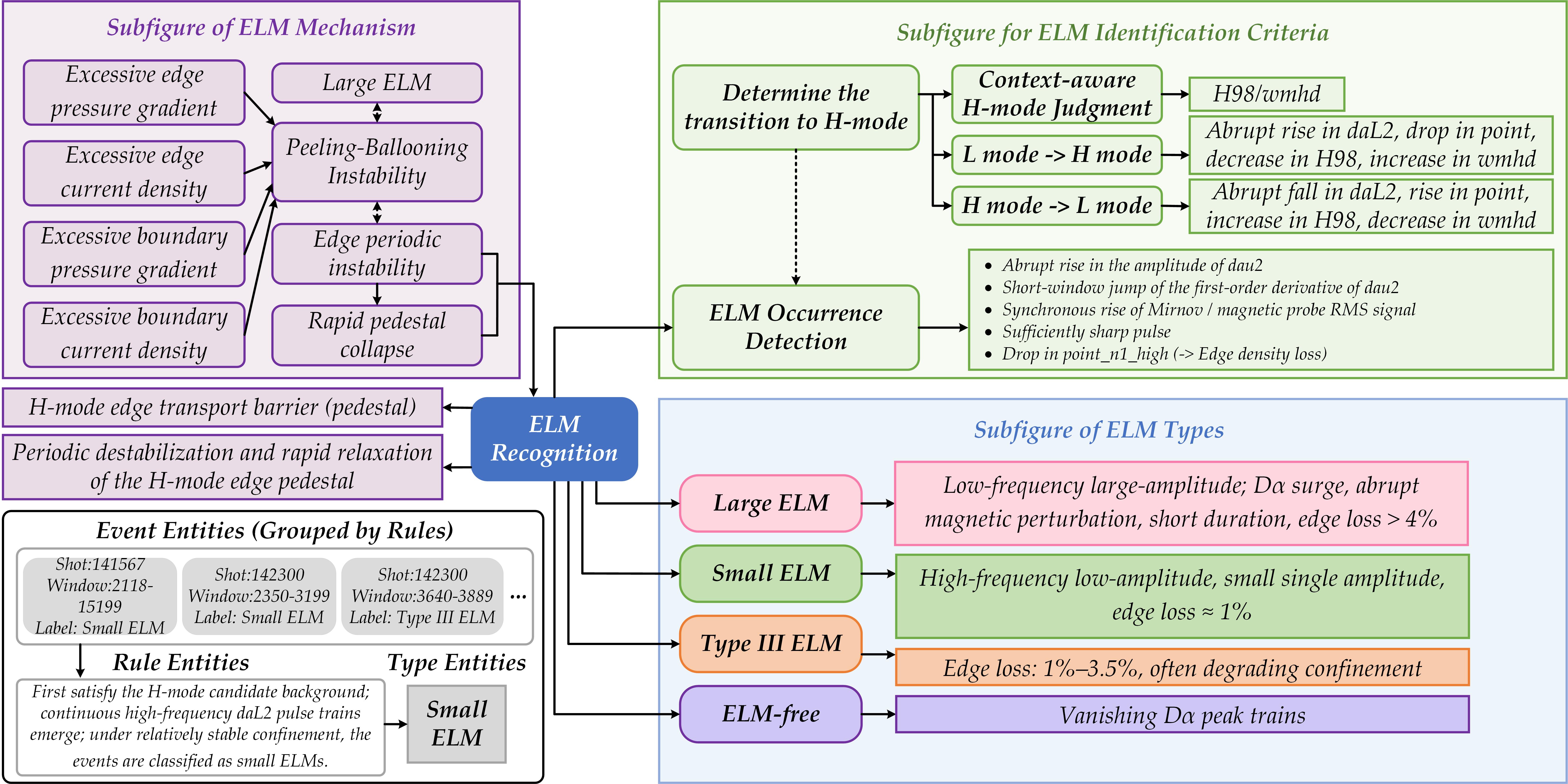}
\caption{An overview of our proposed ELM knowledge resource.}
\label{fig:kg_visualization}
\end{figure*}

Fig.~\ref{fig:kg_visualization} presents the knowledge resource in a
more interpretable semantic view. It groups domain concepts into four
connected aspects: ELM principle, recognition, classification, and
event descriptions. The principle subgraph describes physical causes
and characteristic plasma behavior; the recognition subgraph links
observable diagnostics to ELM evidence; the classification subgraph
organizes distinctions among the four target states; and the event
subgraph records event-related information used to contextualize
historical cases. This figure is therefore a semantic visualization of
the resource rather than a literal drawing of the 30-node computational
graph used by the R-GCN.

\noindent\textbf{Historical memory and knowledge isolation.}
Relational knowledge captures general physical structure, but it does
not represent the full variability of real EAST discharges. We
therefore complement the graph with a case memory constructed from a
fixed subset of 100 training discharges. After extracting structured
candidate windows and applying role and quality filtering, the source
corpus contains 28,688 structured records and retains 6,493
representative cases for retrieval. Each case stores its physics-aware
descriptor together with prototype information, class-support
statistics, quality/safety attributes, and source metadata. At
inference time, the current descriptor can thus be compared with
historical patterns that were actually observed rather than relying on
symbolic knowledge alone.

To prevent information leakage, knowledge construction is isolated at
the discharge level. Let $\mathcal{D}_{\mathrm{tr}}$,
$\mathcal{D}_{\mathrm{val}}$, and $\mathcal{D}_{\mathrm{te}}$ denote
the training, validation, and test discharge sets, and let
$\mathcal{D}_{\mathrm{KB}}$ be the 100-discharge construction subset.
We enforce
\begin{equation}
    \mathcal{D}_{\mathrm{KB}}\subset\mathcal{D}_{\mathrm{tr}},
    \qquad
    \mathcal{D}_{\mathrm{KB}}\cap
    (\mathcal{D}_{\mathrm{val}}\cup\mathcal{D}_{\mathrm{te}})
    =\varnothing.
\label{eq:knowledge_isolation}
\end{equation}
The construction discharges remain part of ordinary supervised
training rather than forming an extra evaluation set. During training,
retrieval additionally excludes cases originating from the query
discharge, preventing nearly identical same-shot windows from serving
as trivial neighbors. Validation and test queries retrieve only from
the training-built memory.

\subsection{Knowledge Graph Collaborative Fusion Module}
\label{subsec:kgcfm}

As illustrated in Fig.~\ref{fig:kgcfm_framework}, KGCFM transforms
the above resource into sample-specific knowledge signals through
three pathways: an online physics pathway, a relational
graph pathway, and a historical retrieval pathway. The three sources
are kept separate initially because they have different resolutions
and failure modes. Physics evidence can provide fine temporal cues but
may be noisy; the graph provides stable relations but is relatively
coarse; retrieval supplies empirical precedents but depends on neighbor
quality. KGCFM therefore extracts both representations and reliability
signals before combining them.

\noindent\textbf{Online physics pathway.}
For every query window, the physics analyzer produces the descriptor
$\boldsymbol{q}$, a physics reliability score $r_{\mathrm{phy}}$, and
an eight-channel local physics sequence
$\boldsymbol{S}_{\mathrm{phy}}\in\mathbb{R}^{B\times T\times 8}$.
The descriptor summarizes window-level evidence, whereas
$\boldsymbol{S}_{\mathrm{phy}}$ retains where the physical responses
occur inside the window. A lightweight temporal encoder maps this
sequence to a hidden physics representation and aligns it with the
token length of the chosen backbone:
\begin{equation}
\begin{aligned}
    \boldsymbol{H}_{\mathrm{phy}}
    &=E_{\mathrm{phy}}(\boldsymbol{S}_{\mathrm{phy}}),\\
    \Delta\boldsymbol{H}_{p}
    &=\operatorname{LN}\!\left(
      \boldsymbol{W}_{p}\operatorname{Align}
      (\boldsymbol{H}_{\mathrm{phy}},N)\right),\\
    \boldsymbol{p}_{\mathrm{phy}}
    &=\operatorname{Softmax}(\boldsymbol{z}_{\mathrm{phy}}).
\end{aligned}
\label{eq:physics_outputs}
\end{equation}
Here, $\Delta\boldsymbol{H}_{p}\in\mathbb{R}^{B\times N\times d}$
is a time-varying physics residual, while
$\boldsymbol{z}_{\mathrm{phy}}$ is produced by a separate physics
decision encoder from the window-level evidence. Adaptive alignment
allows the same pathway to cooperate with backbones using different
tokenizations. Keeping the temporal residual and decision prior
distinct also prevents a single window-level physics score from
overwriting local temporal structure.

\noindent\textbf{Relational graph pathway.}
For the typed graph, relation-specific message transformations are
used to distinguish supporting, contradictory, indicative, and
associative links. A two-layer R-GCN updates node representations as
\begin{equation}
    \boldsymbol{h}_{i}^{(l+1)}
    =\sigma\!\left(
      \boldsymbol{W}_{0}^{(l)}\boldsymbol{h}_{i}^{(l)}
      +\sum_{r\in\mathcal{R}}
       \sum_{j\in\mathcal{N}_{i}^{r}}
       \alpha_{ij}^{r}
       \boldsymbol{W}_{r}^{(l)}\boldsymbol{h}_{j}^{(l)}
      \right),
\label{eq:rgcn_update}
\end{equation}
where $\mathcal{R}$ is the relation set and $\alpha_{ij}^{r}$ includes
the predefined edge strength and normalization. Although the graph is
shared across samples, the pooled graph representation must reflect
the current plasma state. We therefore derive node activation weights
from $\boldsymbol{q}$ and obtain a query-conditioned graph token
\begin{equation}
    \boldsymbol{z}_{G}
    =\operatorname{Pool}_{\boldsymbol{q}}
      (\operatorname{RGCN}^{(2)}(\mathcal{G})).
\label{eq:graph_pooling}
\end{equation}
Thus, the graph topology provides stable relational structure, while
the query descriptor determines which parts of that structure are most
relevant to the current window.

\noindent\textbf{Historical retrieval pathway.}
The current descriptor also queries the training-only memory. Initial
neighbors are selected according to descriptor similarity, after which
their contributions are recalibrated using prototype consistency,
agreement with the physical prior, case quality, and safety cues.
Denoting the selected Top-$K_r$ encoded cases by
$\boldsymbol{M}_{K_r}$, the retrieval operator can be summarized as
\begin{equation}
\begin{aligned}
    (\boldsymbol{M}_{K_r},\widetilde{\boldsymbol{p}}_{\mathrm{ret}})
    &=\mathcal{R}_{\mathrm{case}}
      (\boldsymbol{q},\mathcal{M}_{\mathrm{case}},
       \mathcal{I}_{\mathrm{ret}}),\\
    \boldsymbol{p}_{\mathrm{ret}}
    &=\operatorname{Softmax}(\boldsymbol{z}_{\mathrm{ret}}).
\end{aligned}
\label{eq:case_retrieval}
\end{equation}
Here, $\widetilde{\boldsymbol{p}}_{\mathrm{ret}}$ denotes the
similarity-reweighted class distribution obtained directly from the
retrieved cases.
A retrieval decision encoder aggregates the reweighted evidence to
produce a representation $\boldsymbol{h}_{\mathrm{ret}}$, logits
$\boldsymbol{z}_{\mathrm{ret}}$, and a reliability score
$r_{\mathrm{ret}}$. Unsafe candidates are masked before they can enter
feature-level knowledge attention. Consequently, retrieval is not a
simple nearest-neighbor vote: historical cases provide learnable
context, while their reliability and consistency determine how much
they should affect the current prediction.

The graph token and the retained case tokens are concatenated as
\begin{equation}
    \boldsymbol{K}_{\mathrm{ctx}}
    =[\boldsymbol{z}_{G};\boldsymbol{M}_{K_r}],
\label{eq:graph_case_context}
\end{equation}
which forms the context used later by cross-attention. Notice that the
physics sequence is not included in $\boldsymbol{K}_{\mathrm{ctx}}$;
it follows its own residual path so that continuous local physics and
discrete graph--case knowledge remain distinguishable.

\noindent\textbf{Cross-source knowledge coordination.}
Physics and retrieval may support the same state, but they may also
disagree in atypical windows. We quantify their distributional
agreement using the Jensen--Shannon divergence,
\begin{equation}
    c_{\mathrm{pr}}
    =1-\frac{\operatorname{JS}
      (\boldsymbol{p}_{\mathrm{phy}},
       \boldsymbol{p}_{\mathrm{ret}})}{\log 2},
\label{eq:prior_agreement}
\end{equation}
so $c_{\mathrm{pr}}$ increases when the two knowledge priors are
consistent. A coordinator jointly considers the physics and retrieval
representations, their reliability estimates, retrieval quality, and
$c_{\mathrm{pr}}$ to produce a coordinated representation
$\boldsymbol{h}_{K}$, an adaptive mixture coefficient $\eta$, and a
unified reliability $r_K$. The final knowledge logits are
\begin{equation}
\begin{aligned}
    \boldsymbol{z}_{K}
    &=\eta\boldsymbol{z}_{\mathrm{phy}}
      +(1-\eta)\boldsymbol{z}_{\mathrm{ret}}
      +\boldsymbol{W}_{I}\boldsymbol{h}_{K},\\
    \boldsymbol{p}_{K}
    &=\operatorname{Softmax}(\boldsymbol{z}_{K}).
\end{aligned}
\label{eq:knowledge_coordination}
\end{equation}
This formulation avoids treating every knowledge source as equally
trustworthy. The first two terms interpolate between physics and
historical evidence, while the interaction term allows the coordinator
to model information that becomes useful only when the sources are
considered jointly.

\subsection{Reliability-Aware Dual-Level Fusion}
\label{subsec:dual_level_fusion}

The outputs of KGCFM are incorporated at two complementary levels.
Feature-level fusion supplies contextual information before the shared
prediction head, allowing knowledge to refine hidden temporal
representations. Decision-level fusion operates after classification
and is deliberately restricted to a bounded class correction. This
division gives knowledge enough capacity to resolve ambiguous cases
without replacing the temporal backbone as the primary predictor.

\noindent\textbf{Feature-level fusion.}
Temporal tokens query only the graph--case context through
multi-head cross-attention:
\begin{equation}
    \Delta\boldsymbol{H}_{K}
    =\operatorname{LN}\!\left(
      \boldsymbol{W}_{K}\operatorname{MHA}
      (\boldsymbol{H}_{T},\boldsymbol{K}_{\mathrm{ctx}},
       \boldsymbol{K}_{\mathrm{ctx}})\right).
\label{eq:knowledge_attention}
\end{equation}
Hence, $\boldsymbol{Q}=\boldsymbol{H}_{T}$ and
$\boldsymbol{K}=\boldsymbol{V}=\boldsymbol{K}_{\mathrm{ctx}}$.
This lets each temporal token select relevant relational or historical
context rather than broadcasting the same case information everywhere.
The physics residual follows an independent aligned residual path. The
two signals are combined as
\begin{equation}
\begin{aligned}
    \boldsymbol{H}_{\mathrm{fused}}
    &=\boldsymbol{H}_{T}
      +s_{\mathrm{phy}}g_{\mathrm{phy}}^{f}
       \Delta\boldsymbol{H}_{p}
      +s_{K}g_{K}^{f}\Delta\boldsymbol{H}_{K},\\
    \boldsymbol{Z}_{\mathrm{feat}}
    &=h_{\psi}(\boldsymbol{H}_{\mathrm{fused}},\mathcal{C}_{T}).
\end{aligned}
\label{eq:feature_level_fusion}
\end{equation}
The gates $g_{\mathrm{phy}}^{f}$ and $g_{K}^{f}$ depend on base
uncertainty, source reliability, and retrieval quality. The residual
strengths $s_{\mathrm{phy}}$ and $s_K$ are bounded and initialized
conservatively, so the model begins close to the temporal baseline and
learns to inject knowledge only when it improves the supervised
objective. This design is particularly useful when a physically
meaningful cue is present but the corresponding retrieved cases are
weak, or vice versa, because the two sources need not be amplified
together.

\noindent\textbf{Decision-level correction.}
Feature fusion alone does not explicitly guarantee that knowledge
influence decreases when the data-driven prediction is already
decisive. We therefore introduce a separate reliability-aware decision
gate,
\begin{equation}
    \Delta\boldsymbol{z}_{\mathrm{KG}}
    =G_{\mathrm{dec}}
    (\boldsymbol{z}_{K},r_{\mathrm{phy}},r_{\mathrm{ret}},r_{K},
     c_{\mathrm{base}},u_{\mathrm{base}}),
\label{eq:decision_correction}
\end{equation}
which converts coordinated knowledge logits into a bounded
window-level correction. High knowledge reliability together with
large base uncertainty permits stronger assistance; disagreement,
low retrieval reliability, or a confident base prediction suppresses
the correction. Therefore, the knowledge pathway behaves as conditional
support rather than an unconditional second classifier.

Class imbalance requires an additional safeguard because an incorrect
correction can disproportionately erase rare-state predictions. Let
$\mathcal{Y}_{m}=\{2,3\}$ denote Type-III and Large ELM, and define
$p_b(\mathcal{Y}_{m})=\sum_{k\in\mathcal{Y}_{m}}p_{b,k}$. We detect a
potentially harmful correction when the base branch already supports a
minority state but the coordinated knowledge distribution reduces that
support by more than a margin $m$:
\begin{equation}
\begin{aligned}
    \chi_b
    &=\mathbb{I}\!\left[
      \widehat{y}^{\,\mathrm{base}}_b\in\mathcal{Y}_{m}
      \ \mathrm{or}\ 
      p_{\mathrm{base},b}(\mathcal{Y}_{m})\ge\tau\right]\\
    &\quad\cdot\mathbb{I}\!\left[
      p_{K,b}(\mathcal{Y}_{m})+m
      <p_{\mathrm{base},b}(\mathcal{Y}_{m})\right],\\
    \gamma_b
    &=1-(1-\gamma_{\mathrm{protect}})\chi_b,\\
    \widetilde{\Delta\boldsymbol{z}}_{\mathrm{KG},b}
    &=\gamma_b\Delta\boldsymbol{z}_{\mathrm{KG},b}.
\end{aligned}
\label{eq:minority_protection}
\end{equation}
The mechanism does not force a minority prediction; it only attenuates
knowledge correction when that correction is likely to remove a
reasonable minority hypothesis from the temporal model.

Finally, the protected decision residual is broadcast over the window
and added to the point-wise feature-fused logits:
\begin{equation}
\begin{aligned}
    \boldsymbol{Z}_{\mathrm{final}}
    &=\boldsymbol{Z}_{\mathrm{feat}}
      +\lambda_{\mathrm{DS}}\operatorname{Broadcast}
      (\widetilde{\Delta\boldsymbol{z}}_{\mathrm{KG}}),\\
    \boldsymbol{P}_{\mathrm{final}}
    &=\operatorname{Softmax}(\boldsymbol{Z}_{\mathrm{final}}).
\end{aligned}
\label{eq:final_prediction}
\end{equation}
Although the decision correction is shared across positions within a
window, it is only an additive class bias. The temporal variation of
the final output still comes from $\boldsymbol{Z}_{\mathrm{feat}}$.
Thus, the model retains point-wise temporal resolution while using
window-level knowledge to adjust the overall decision tendency.

\subsection{Optimization Objective}
\label{subsec:optimization_objective}

We train the temporal and knowledge branches jointly using
\begin{equation}
    \mathcal{L}
    =\mathcal{L}_{\mathrm{cls}}
    +\mathcal{L}_{\mathrm{know}}
    +\mathcal{L}_{\mathrm{safe}}.
\label{eq:overall_objective}
\end{equation}
The primary classification objective operates on all point-wise
positions and combines complementary treatments of imbalance:
\begin{equation}
    \mathcal{L}_{\mathrm{cls}}
    =\mathcal{L}_{\mathrm{CBCE}}
    +0.30\mathcal{L}_{\mathrm{focal}}
    +0.20\mathcal{L}_{\mathrm{softF1}}.
\label{eq:classification_loss}
\end{equation}
$\mathcal{L}_{\mathrm{CBCE}}$ uses class weights computed from the
training split, the focal term with exponent $1.5$ emphasizes difficult
or underrepresented positions, and the differentiable Soft-F1 term
encourages balanced class-wise performance. These losses are evaluated
after flattening the batch and time dimensions so that every point
contributes to the recognition objective.

The knowledge branch is trained to remain predictive and internally
consistent rather than relying only on gradients propagated through
the final logits. We group its auxiliary objectives as
\begin{equation}
\begin{aligned}
    \mathcal{L}_{\mathrm{know}}
    &=\lambda_{\mathrm{prior}}\mathcal{L}_{\mathrm{prior}}
      +\lambda_{\mathrm{aux}}\mathcal{L}_{\mathrm{aux}}
      +\lambda_G\mathcal{L}_G
      +\lambda_{\mathrm{cons}}\mathcal{L}_{\mathrm{cons}}.
\end{aligned}
\label{eq:knowledge_loss}
\end{equation}
Here, the prior and auxiliary terms supervise the physics, retrieval,
and coordinated knowledge predictions at the window level;
$\mathcal{L}_G$ maintains discriminative graph representations; and
$\mathcal{L}_{\mathrm{cons}}$ encourages compatible sources to agree
without requiring identical predictions. Window-majority labels are
used only for these window-level auxiliary objectives and never replace
the point-wise labels in $\mathcal{L}_{\mathrm{cls}}$.

Finally, safe-fusion regularization explicitly constrains how knowledge
modifies the backbone:
\begin{equation}
\begin{aligned}
    \mathcal{L}_{\mathrm{safe}}
    &=\lambda_{\mathrm{help}}\mathcal{L}_{\mathrm{help}}
      +\lambda_{\mathrm{minor}}\mathcal{L}_{\mathrm{minor}}
      +\lambda_{\mathrm{pres}}\mathcal{L}_{\mathrm{pres}}
      +\lambda_{\mathrm{gate}}\mathcal{L}_{\mathrm{gate}}.
\end{aligned}
\label{eq:safe_loss}
\end{equation}
$\mathcal{L}_{\mathrm{help}}$ rewards knowledge corrections that improve
the supervised decision, whereas $\mathcal{L}_{\mathrm{minor}}$
penalizes harmful suppression of minority classes.
$\mathcal{L}_{\mathrm{pres}}$ discourages unnecessary distortion of
temporal features when knowledge is unreliable, and
$\mathcal{L}_{\mathrm{gate}}$ regularizes the adaptive gates to avoid
persistent saturation. Together, the three loss groups train KGF-ELM
to improve recognition using knowledge while preserving the robustness
of the data-driven temporal predictor.

\section{Experiments}
\label{sec:experiments}

\subsection{Experimental Setup}
\label{subsec:experimental_setup}

\noindent\textbf{Dataset and evaluation protocol.}
Experiments are conducted on EAST-ELM640~\cite{si2025hgts}, which
contains 640 plasma discharges and 18 diagnostic and plasma-state
signals. The four point-wise labels are L-mode, Small/No ELM,
Type-III ELM, and Large ELM. Because neighboring windows from the same
discharge are strongly correlated, all data isolation is performed at
the discharge level rather than by randomly splitting windows. The
dataset contains 448 training, 96 validation, and 96 test discharges,
and these three sets remain mutually exclusive throughout model
training, checkpoint selection, knowledge construction, and final
evaluation.

A fixed subset of 100 training discharges is additionally used for
offline knowledge-resource construction and graph-schema verification. Structured
records from these discharges are used to validate graph concepts and
relations, while the same training-only subset provides the historical
case memory and retrieval index. These discharges are
not an additional evaluation subset: they remain part of the 448-shot
supervised training set. No validation or test discharge contributes
to any data-dependent knowledge component. During training, retrieval
also excludes historical cases originating from the current query
discharge, while validation and test queries can access training cases
only. This protocol is important because the stride-one segmentation
produces many overlapping windows; without discharge-level isolation,
near-duplicate local patterns could otherwise appear on both sides of
the evaluation boundary.

The continuous signals are segmented using windows of length $T=100$
and stride one. Each sample is therefore represented as
$\boldsymbol{X}\in\mathbb{R}^{B\times100\times18}$ and produces
point-wise logits in $\mathbb{R}^{B\times100\times4}$. The primary
classification loss and reported metrics are evaluated after
flattening the batch and temporal dimensions, so every temporal
position contributes to evaluation. Window-majority labels are used
only for the auxiliary window-level knowledge objectives and never
replace the point-wise supervision.

\noindent\textbf{Temporal backbones and controlled comparison.}
To test whether KGCFM depends on a specific temporal representation,
we integrate it with five backbones: Timer-XL~\cite{liu2025timer},
CrossLinear~\cite{zhou2025crosslinear}, Timer~\cite{liu2024timer},
TimeXer~\cite{wang2024timexer}, and TimeFilter~\cite{hu2025timefilter}.
CrossLinear equipped with the complete KGCFM is used as the default
KGF-ELM in the broad model comparison. For the controlled generality
experiment, only the temporal encoder is changed; the data split,
input windows, knowledge resource, retrieval protocol, loss,
knowledge-fusion module, warm-up schedule, and evaluation procedure
remain fixed. Each enhanced model is therefore compared directly with
its own temporal-only counterpart.

The patch length is fixed to 20 for the five controlled backbones.
Where the corresponding architectural parameter is applicable, the
hidden dimension, feed-forward dimension, number of encoder layers,
number of attention heads, and dropout are set to 256, 512, 2, 8,
and 0.1, respectively. Backbone-specific operations are retained
rather than forcing all encoders into the same internal design. A
common feature interface exposes the hidden temporal tokens to KGCFM.
When token lengths differ, the local physics sequence is adaptively
aligned to the backbone representation and graph--case tokens are
projected to the required hidden dimension. Thus, the knowledge
resource and fusion mechanism remain unchanged while the temporal
representation varies.

\noindent\textbf{Training configuration.}
All models are optimized using AdamW with batch size 192 and an
initial learning rate of $10^{-4}$. We use cosine annealing, zero
weight decay, and gradient-norm clipping at 1.0. Training proceeds for
at most 10 epochs. Validation Macro-F1 is evaluated after each epoch,
early stopping uses patience 3, and the checkpoint with the highest
validation Macro-F1 is restored for a single final test evaluation.
The primary experiments use random seed 2026. Unless otherwise
specified, the historical retrieval size is $K_r=5$ and the overall
decision-fusion scale is $\lambda_{\mathrm{DS}}=0.50$.

Knowledge fusion is introduced conservatively during optimization.
The feature-fusion pathway is present from the beginning, but its
physics and knowledge residual strengths are zero-initialized. The
decision-level gates are held at zero during the first two epochs and
released from epoch 3. The initial optimization is therefore close to
the temporal predictor, after which feature enhancement and decision
correction are learned progressively. Experiments are implemented in
PyTorch 2.5.1 with CUDA 12.1 on NVIDIA GeForce RTX 4090 GPUs.

\noindent\textbf{Evaluation metrics.}
We report Accuracy, Macro-Precision, Macro-Recall, Macro-F1, and
one-vs-rest Macro-AUC. EAST-ELM640 is strongly imbalanced, with
Type-III and especially Large ELM occurring much less frequently than
the dominant Small/No ELM class. Macro-F1 is therefore treated as the
primary comparison and checkpoint-selection metric because it gives
equal importance to all four categories. Macro-Recall measures class
coverage, Macro-Precision reflects the reliability of predicted
classes, and Macro-AUC characterizes class separability independently
of a single operating threshold. The combination of these metrics is
used throughout the following analyses rather than relying on overall
Accuracy alone.

\subsection{Overall Recognition Performance}
\label{subsec:overall_performance}

We evaluate the proposed framework from two complementary
perspectives. First, KGF-ELM is compared with a broader set of
representative temporal models to establish its overall performance.
Second, five backbones are evaluated before and after integrating the
same KGCFM to determine whether the observed gain is transferable
across temporal architectures. Tables~\ref{tab:main_comparison} and
\ref{tab:backbone_comparison} report the two comparisons separately.

\begin{table}[!t]
\centering
\caption{Comparison with representative temporal models on
EAST-ELM640 (\%). Best and second-best results are shown in bold and
underlined.}
\label{tab:main_comparison}
\scriptsize
\setlength{\tabcolsep}{1.7pt}
\renewcommand{\arraystretch}{1.08}
\begin{tabular*}{\columnwidth}{@{\extracolsep{\fill}}lccccc@{}}
\toprule
Model & Acc. & P & R & F1 & AUC \\
\midrule
KGF-ELM
& \textbf{91.7} & \textbf{82.9} & \underline{80.5}
& \textbf{81.6} & \textbf{98.0} \\
\addlinespace[2pt]
Timer-XL & 88.6 & 78.8 & 76.8 & 77.3 & 96.5 \\
TimeFilter & 86.0 & 72.9 & 76.6 & 74.7 & 95.0 \\
CrossLinear
& \underline{90.1} & \underline{80.2} & 79.2
& \underline{79.2} & \underline{97.5} \\
Timer & 82.9 & 73.4 & 64.0 & 68.0 & 92.0 \\
iTransformer & 84.7 & 72.3 & 70.3 & 70.8 & 94.5 \\
TimeMixer & 85.5 & 80.0 & 52.7 & 56.9 & 86.9 \\
TimeXer & 88.0 & 75.4 & \textbf{81.7} & 77.8 & 96.4 \\
SOFTS & 86.7 & 77.6 & 69.5 & 72.7 & 95.2 \\
PatchTST & 85.8 & 75.5 & 72.9 & 73.2 & 94.9 \\
\bottomrule
\end{tabular*}
\vspace{3pt}
\end{table}

\noindent\textbf{Comparison with representative temporal models.}
As shown in Table~\ref{tab:main_comparison}, KGF-ELM achieves the best Accuracy,
Macro-Precision, Macro-F1, and Macro-AUC, reaching $91.7\%$,
$82.9\%$, $81.6\%$, and $98.0\%$, respectively. CrossLinear is the
strongest temporal-only model in terms of Accuracy, Macro-F1, and
Macro-AUC, with $90.1\%$, $79.2\%$, and $97.5\%$, while TimeXer
achieves the highest baseline Macro-Recall of $81.7\%$. KGF-ELM thus
does not obtain its advantage simply because the selected temporal
backbone is weak: it improves upon the already competitive
CrossLinear representation and provides the best overall balance
among the reported aggregate metrics.

The differences among temporal baselines also illustrate why a
single metric is insufficient in this task. TimeMixer, for example,
has relatively high Macro-Precision but substantially lower
Macro-Recall, whereas TimeXer emphasizes recall more strongly. Such
variation indicates that temporal architectures learn different
decision tendencies under class imbalance. KGF-ELM combines high
precision with competitive recall, leading to the highest Macro-F1.
Its Macro-AUC of $98.0\%$ further shows that the improvement is not
limited to the final argmax decisions but is accompanied by strong
overall class separability.

\begin{table}[!t]
\centering
\caption{Performance before and after integrating KGCFM (\%). Bold indicates the better result within each backbone.}
\label{tab:backbone_comparison}
\scriptsize
\setlength{\tabcolsep}{1.5pt}
\renewcommand{\arraystretch}{1.08}
\begin{tabular*}{\columnwidth}{@{\extracolsep{\fill}}llccccc@{}}
\toprule
Backbone & Variant & Acc. & P & R & F1 & AUC \\
\midrule
\multirow{2}{*}{Timer-XL}
& Base & 88.6 & \textbf{78.8} & 76.8 & 77.3 & 96.5 \\
& +KGCFM & \textbf{90.9} & 77.4 & \textbf{83.4} & \textbf{79.8} & \textbf{97.6} \\
\addlinespace[2pt]
\multirow{2}{*}{CrossLinear}
& Base & 90.1 & 80.2 & 79.2 & 79.2 & 97.5 \\
& +KGCFM & \textbf{91.7} & \textbf{82.9} & \textbf{80.5} & \textbf{81.6} & \textbf{98.0} \\
\addlinespace[2pt]
\multirow{2}{*}{Timer}
& Base & 82.9 & 73.4 & 64.0 & 68.0 & 92.0 \\
& +KGCFM & \textbf{89.9} & \textbf{77.6} & \textbf{79.5} & \textbf{78.4} & \textbf{96.9} \\
\addlinespace[2pt]
\multirow{2}{*}{TimeXer}
& Base & 88.0 & 75.4 & 81.7 & 77.8 & 96.4 \\
& +KGCFM & \textbf{90.5} & \textbf{78.2} & \textbf{84.5} & \textbf{80.8} & \textbf{97.6} \\
\addlinespace[2pt]
\multirow{2}{*}{TimeFilter}
& Base & 86.0 & 72.9 & 76.6 & 74.7 & 95.0 \\
& +KGCFM & \textbf{91.1} & \textbf{78.4} & \textbf{82.1} & \textbf{80.1} & \textbf{97.3} \\
\bottomrule
\end{tabular*}
\vspace{3pt}
\end{table}

\noindent\textbf{Backbone generality.}
Table~\ref{tab:backbone_comparison} provides a stricter paired comparison. After integrating
KGCFM, Macro-F1 improves by $2.5$, $2.4$, $10.4$, $3.0$, and $5.4$
percentage points for Timer-XL, CrossLinear, Timer, TimeXer, and
TimeFilter, respectively, corresponding to an average gain of
$4.74$ points. Accuracy, Macro-Recall, and Macro-AUC improve for all
five backbones. Macro-Precision also increases for four of the five
models, showing that the recall gains are generally not obtained by
indiscriminately increasing positive predictions.

The magnitude of improvement depends on the information already
captured by the temporal encoder. Timer exhibits the largest gain:
its Macro-Recall increases from $64.0\%$ to $79.5\%$ and Macro-F1
from $68.0\%$ to $78.4\%$. TimeFilter also gains $5.4$ Macro-F1
points, while TimeXer, despite starting from the strongest baseline
recall, further improves Recall from $81.7\%$ to $84.5\%$ and F1
from $77.8\%$ to $80.8\%$. These results suggest that knowledge can
both compensate for missing cues in weaker representations and refine
already competitive temporal models.

Looking across the five paired experiments, the average improvement
is $3.70$ percentage points in Accuracy, $6.34$ points in
Macro-Recall, $4.74$ points in Macro-F1, and $2.00$ points in
Macro-AUC. The larger average gain in Macro-Recall than in Accuracy is
particularly relevant under class imbalance: the knowledge pathway
does more than refine predictions for already dominant states and
instead improves coverage of classes that are more easily missed by a
pure temporal model. At the same time, the consistent AUC gains imply
that this change is accompanied by improved ranking of class
probabilities rather than only a shift in the final decision boundary.

CrossLinear provides the strongest full configuration, improving from
$79.2\%$ to $81.6\%$ Macro-F1 while simultaneously increasing all
other reported metrics. Timer-XL shows a different but informative
trade-off: its Macro-Precision decreases from $78.8\%$ to $77.4\%$,
whereas Macro-Recall increases from $76.8\%$ to $83.4\%$. The net
Macro-F1 improvement indicates that additional knowledge enables the
model to recover more difficult ELM positions at a modest precision
cost. Overall, the consistent paired gains support the intended role
of KGCFM as a transferable knowledge-enhancement module rather than a
backbone-specific classifier.

\subsection{Ablation Study}
\label{subsec:ablation}

We conduct two levels of ablation using Timer-XL and CrossLinear to
separate architectural effects from the contributions of individual
knowledge sources. Table~\ref{tab:overall_ablation} examines feature
fusion, decision fusion, and reliability control at the framework
level. Table~\ref{tab:kgcfm_ablation} removes physical evidence,
prototype reasoning, graph encoding, and historical retrieval
individually from KGCFM. All variants use the same data split and
evaluation protocol as the corresponding full model.

\begin{table}[!t]
\centering
\caption{Overall module ablation in Macro-F1 (\%).}
\label{tab:overall_ablation}
\small
\setlength{\tabcolsep}{6pt}
\renewcommand{\arraystretch}{1.10}
\begin{tabular*}{\columnwidth}{@{\extracolsep{\fill}}lcc@{}}
\toprule
Variant & Timer-XL & CrossLinear \\
\midrule
w/o Knowledge Module & 77.3 & 79.2 \\
w/o Feature Fusion & 78.1 & 79.3 \\
w/o Decision Fusion & 79.5 & 80.1 \\
w/o Reliability & 79.6 & 81.2 \\
\addlinespace[2pt]
Full Model & \textbf{79.8} & \textbf{81.6} \\
\bottomrule
\end{tabular*}
\end{table}

\noindent\textbf{Overall fusion architecture.}
Removing the complete knowledge module reduces Macro-F1 by $2.5$
points for Timer-XL and $2.4$ points for CrossLinear, recovering the
corresponding temporal-only performance. Among the remaining variants,
feature-level fusion has the largest effect. Without this pathway,
Macro-F1 falls from $79.8\%$ to $78.1\%$ for Timer-XL and from
$81.6\%$ to $79.3\%$ for CrossLinear. The $1.7$/$2.3$-point drops
indicate that incorporating knowledge before the classification head
is the primary source of the enhancement.

This result is consistent with the design of the feature pathway.
The time-varying physics residual can directly complement the temporal
representation, while graph and historical-case tokens supply
query-dependent semantic context through cross-attention. Removing
feature fusion leaves the knowledge module with only decision-level
class correction, which cannot recover the same representation-level
benefit. In contrast, removing decision fusion produces smaller drops
of $0.3$ and $1.5$ points. Decision correction is therefore useful but
acts mainly as a second-stage refinement after knowledge-aware temporal
features have already been constructed.

Reliability modeling provides a third, different role. Removing it
decreases Macro-F1 by $0.2$ points on Timer-XL and $0.4$ points on
CrossLinear. The numerical effect is smaller because reliability does
not introduce an additional knowledge source. Instead, it determines
how much the existing sources should affect a given query. Its
consistent positive contribution supports the use of adaptive gating
to suppress uncertain or conflicting knowledge rather than applying a
fixed fusion strength to every window. Taken together, the overall
ablation supports a hierarchical design: feature enhancement provides
the main gain, decision fusion adds complementary correction, and
reliability estimation constrains when these corrections should be
trusted.

\begin{table}[!t]
\centering
\caption{KGCFM component ablation in Macro-F1 (\%).}
\label{tab:kgcfm_ablation}
\small
\setlength{\tabcolsep}{6pt}
\renewcommand{\arraystretch}{1.10}
\begin{tabular*}{\columnwidth}{@{\extracolsep{\fill}}lcc@{}}
\toprule
Variant & Timer-XL & CrossLinear \\
\midrule
w/o Physical Evidence & 78.7 & 79.7 \\
w/o Prototype Reasoning & 79.3 & 80.8 \\
w/o Knowledge Graph & 79.2 & 81.2 \\
w/o Historical Retrieval & 78.9 & 79.6 \\
\addlinespace[2pt]
Full KGCFM & \textbf{79.8} & \textbf{81.6} \\
\bottomrule
\end{tabular*}
\end{table}

\noindent\textbf{Internal knowledge components.}
Physical evidence and historical retrieval are the most influential
internal sources. Removing the physical pathway decreases Macro-F1
by $1.1$ points for Timer-XL and $1.9$ for CrossLinear. The current
window therefore contains physics-aware information that is not fully
captured by the learned temporal features. This contribution includes
both the time-varying physics residual and window-level physical
support used during knowledge coordination.

Historical retrieval produces a similarly strong effect. Removing it
reduces Macro-F1 by $0.9$ points for Timer-XL and $2.0$ points for
CrossLinear, the largest internal drop for the latter backbone. The
result indicates that similar training discharges provide useful
instance-level evidence even when the temporal encoder itself is
strong. Importantly, this information comes from the training-only
memory and is therefore historical support rather than access to
validation or test examples.

Prototype reasoning and the relational graph yield smaller but
consistent improvements. Removing prototypes causes $0.5$/$0.8$-point
drops, while removing graph encoding causes $0.6$/$0.4$-point drops
for Timer-XL/CrossLinear. Prototypes provide an intermediate semantic
description between low-level physical evidence and class support,
whereas the graph models dependencies among evidence, prototypes, and
ELM classes. The different magnitudes across the two backbones also
suggest that KGCFM does not apply an identical correction to every
encoder; each temporal representation benefits from the knowledge it
lacks. The complete configuration nevertheless performs best in both
cases, confirming the complementarity of the four sources.

The two ablation levels also clarify why no single knowledge source
should be interpreted as the complete explanation for the improvement.
At the architectural level, the largest loss occurs when knowledge is
prevented from modifying temporal features. At the source level,
however, both physical evidence and retrieved historical experience
remain important, and the prototype and graph pathways provide
additional gains. In other words, the model benefits from both
\emph{what} knowledge is available and \emph{where} that knowledge is
introduced into the temporal predictor. This distinction motivates
the collaborative design of KGCFM rather than replacing it with either
a single auxiliary prior or a simple retrieval vote.

\subsection{Knowledge Resource and Retrieval Analysis}
\label{subsec:knowledge_retrieval_analysis}

Beyond component removal, we examine whether performance is sensitive
to the amount of available knowledge and to two key fusion settings.
Timer-XL with seed 2026 is used as a controlled backbone. Except for
the factor under study, the training, checkpoint-selection, and test
protocol remains unchanged. The default configuration uses a
100-discharge knowledge resource, $K_r=5$, and
$\lambda_{\mathrm{DS}}=0.50$. Each factor is reported in a separate
table to keep the trends visually distinct.

\begin{table}[!t]
\centering
\caption{Knowledge-resource size sensitivity with Timer-XL (\%).}
\label{tab:kb_size_sensitivity}
\scriptsize
\setlength{\tabcolsep}{2.4pt}
\renewcommand{\arraystretch}{1.10}
\begin{tabular*}{\columnwidth}{@{\extracolsep{\fill}}cccccc@{}}
\toprule
KB shots & Acc. & P & R & F1 & AUC \\
\midrule
25  & 88.8 & \textbf{78.7} & 77.5 & 77.6 & 96.6 \\
50  & 89.1 & 78.5 & 78.3 & 77.9 & 96.7 \\
75  & 90.0 & 78.0 & 80.8 & 78.8 & 97.2 \\
\textbf{100} & \textbf{90.9} & 77.4 & \textbf{83.4} & \textbf{79.8} & \textbf{97.6} \\
\bottomrule
\end{tabular*}
\end{table}

\noindent\textbf{Knowledge-resource size.}
Increasing the number of knowledge-construction discharges from 25 to
100 produces a clear upward trend in Accuracy, Macro-Recall,
Macro-F1, and Macro-AUC. Accuracy increases from $88.8\%$ to
$90.9\%$, Macro-Recall from $77.5\%$ to $83.4\%$, Macro-F1 from
$77.6\%$ to $79.8\%$, and Macro-AUC from $96.6\%$ to $97.6\%$.
The improvement is modest from 25 to 50 discharges but becomes more
pronounced as the resource expands to 75 and 100 shots. Relative to
the 25-shot configuration, the 100-shot resource gains $2.2$ Macro-F1
points and $5.9$ Macro-Recall points.

The trend is consistent with the purpose of the knowledge resource.
Additional construction discharges broaden the coverage of confinement
conditions, diagnostic patterns, historical ELM cases, and their
associated prototype-activation patterns. A test query is therefore
more likely to find relevant and diverse training precedents.
Macro-Precision decreases from $78.7\%$ to $77.4\%$, showing that
broader knowledge coverage makes the model slightly more willing to
recover difficult states. However, the simultaneous improvements in
Accuracy, Macro-F1, and Macro-AUC indicate that the increased recall
outweighs this moderate precision cost. The complete 100-discharge
resource is consequently retained as the default.

\begin{table}[!t]
\centering
\caption{Historical retrieval Top-$K_r$ sensitivity with Timer-XL (\%).}
\label{tab:retrieval_topk_sensitivity}
\scriptsize
\setlength{\tabcolsep}{2.4pt}
\renewcommand{\arraystretch}{1.10}
\begin{tabular*}{\columnwidth}{@{\extracolsep{\fill}}cccccc@{}}
\toprule
Top-$K_r$ & Acc. & P & R & F1 & AUC \\
\midrule
1  & 90.6 & 77.1 & 82.9 & 79.4 & 97.4 \\
3  & 90.8 & 77.3 & 83.2 & 79.6 & 97.5 \\
\textbf{5} & \textbf{90.9} & \textbf{77.4} & \textbf{83.4} & \textbf{79.8} & \textbf{97.6} \\
10 & 90.8 & 77.3 & 83.3 & 79.7 & \textbf{97.6} \\
\bottomrule
\end{tabular*}
\end{table}

\noindent\textbf{Historical retrieval size.} 
The retrieval experiment exhibits a different pattern. Performance is
already strong with only the nearest historical case: $K_r=1$ yields
$79.4\%$ Macro-F1 and $82.9\%$ Macro-Recall. Increasing $K_r$ to 3
and then 5 produces small but consistent improvements, reaching
$79.8\%$ Macro-F1 and $83.4\%$ Macro-Recall. The result suggests that
the top-ranked case carries the most direct historical guidance, while
a few additional high-similarity cases contribute complementary
physical or class-level evidence.

Increasing the retrieval size further to $K_r=10$ does not improve
the result: Macro-F1 changes from $79.8\%$ to $79.7\%$, and the other
metrics remain almost unchanged. Lower-ranked candidates are generally
less similar or partly redundant with the highest-ranked cases. Their
effect is additionally reduced by similarity-based reweighting and
retrieval reliability. Consequently, KGCFM remains stable rather than
becoming increasingly sensitive as more candidates are exposed. The
default $K_r=5$ represents a practical balance between historical
coverage and redundant retrieval.

The narrow performance range over the evaluated $K_r$ values is also
useful from a robustness perspective. If recognition quality depended
strongly on an exact neighbor count, the historical pathway would be
sensitive to small changes in case density across plasma regimes. The
observed saturation instead indicates that similarity weighting and
the reliability gate absorb moderate variation in the candidate set.
This behavior is consistent with the intended use of retrieval as a
supporting context rather than a hard nearest-neighbor classifier.

\begin{table}[!t]
\centering
\caption{Decision-fusion scale sensitivity with Timer-XL (\%).}
\label{tab:decision_scale_sensitivity}
\scriptsize
\setlength{\tabcolsep}{2.4pt}
\renewcommand{\arraystretch}{1.10}
\begin{tabular*}{\columnwidth}{@{\extracolsep{\fill}}cccccc@{}}
\toprule
$\lambda_{\mathrm{DS}}$ & Acc. & P & R & F1 & AUC \\
\midrule
0    & 90.1 & \textbf{79.3} & 79.6 & 79.4 & 97.3 \\
0.25 & 90.6 & 77.0 & 83.0 & 79.6 & 97.4 \\
\textbf{0.50} & \textbf{90.9} & 77.4 & \textbf{83.4} & \textbf{79.8} & \textbf{97.6} \\
0.75 & 90.3 & 76.7 & 83.2 & 79.0 & 97.5 \\
\bottomrule
\end{tabular*}
\end{table}

\noindent\textbf{Decision-fusion scale.} 
The final sensitivity study varies the magnitude of the window-level
knowledge correction. When $\lambda_{\mathrm{DS}}=0$, only the
decision-level correction is disabled; feature-level physics and
graph--case fusion remains active. This configuration therefore differs
from the knowledge-free variant in Table~\ref{tab:overall_ablation}.
Increasing the scale from 0 to 0.50 improves Accuracy from $90.1\%$
to $90.9\%$, Macro-Recall from $79.6\%$ to $83.4\%$, and Macro-F1
from $79.4\%$ to $79.8\%$. A scale of 0.25 already provides most of
the recall gain, showing that even a moderate knowledge bias can help
recover ambiguous positions after feature fusion.

The scale should nevertheless remain bounded. At
$\lambda_{\mathrm{DS}}=0.75$, Macro-Recall remains high at $83.2\%$
but Macro-Precision drops to $76.7\%$, reducing Macro-F1 to $79.0\%$.
An overly strong window-level correction can therefore override
locally reliable temporal evidence and shift the precision--recall
balance too far toward recall. The default value 0.50 provides the
best combined Accuracy, Macro-Recall, Macro-F1, and Macro-AUC among
the evaluated settings. Together, the three sensitivity studies show
that KGF-ELM benefits from richer knowledge coverage, but neither more
retrieved cases nor stronger correction is unconditionally better.

These sensitivity experiments use a fixed backbone and seed and are
therefore interpreted as controlled trends rather than estimates of
cross-seed variability. They are intended to characterize the
behavior around the default configuration, while the paired backbone
and ablation experiments provide the primary evidence for the model's
effectiveness.

Taken jointly, the three controlled studies reveal different
sensitivity regimes. Knowledge-resource size has the clearest monotonic
effect because it changes the diversity of available historical
experience. Retrieval size saturates quickly because additional
neighbors become increasingly redundant, whereas decision scale has an
interior optimum because it directly controls how strongly window-level
knowledge can alter point-wise logits. These distinct trends are
consistent with the functional roles assigned to the three mechanisms
in Section~\ref{sec:method} and provide a more informative picture than
a single hyperparameter setting alone.

\subsection{Class-Wise Performance Analysis}
\label{subsec:classwise_analysis}

Aggregate Macro-F1 indicates balanced improvement but does not reveal
which ELM states are most affected. We therefore compare the internal
pre-enhancement prediction and final prediction from the same complete
CrossLinear checkpoint. This is a within-model diagnostic comparison,
not a separately trained knowledge-free baseline; the controlled
contribution of KGCFM is established by the ablations in
Table~\ref{tab:overall_ablation}. Table~\ref{tab:classwise_knowledge_effect}
reports the class-wise changes.

\begin{table*}
\centering
\caption{Class-wise point-wise performance before and after knowledge
enhancement in the complete CrossLinear-based KGF-ELM (\%). Support
denotes the number of evaluated window positions; consecutive windows
overlap because the stride is one.}
\label{tab:classwise_knowledge_effect}
\small
\setlength{\tabcolsep}{5.5pt}
\renewcommand{\arraystretch}{1.08}
\begin{tabular*}{\textwidth}{@{\extracolsep{\fill}}lrrrrrrrr@{}}
\toprule
& & \multicolumn{3}{c}{Before enhancement}
& \multicolumn{3}{c}{Final output} & \\
\cmidrule(lr){3-5}\cmidrule(lr){6-8}
Class & Support & P & R & F1 & P & R & F1 & $\Delta$F1 \\
\midrule
L-mode & 11,883,509 & 52.97 & 67.54 & 59.37
& 84.96 & 85.77 & 85.37 & \textbf{+26.00} \\
Small/No ELM & 61,969,398 & 88.29 & 87.72 & 88.00
& 94.14 & 95.60 & 94.86 & +6.86 \\
Type-III ELM & 7,739,025 & 84.58 & 52.04 & 64.44
& 85.26 & 73.44 & 78.91 & \textbf{+14.47} \\
Large ELM & 1,152,168 & 62.25 & 68.18 & 65.08
& 67.62 & 67.39 & 67.50 & +2.42 \\
\bottomrule
\end{tabular*}
\end{table*}

The support values correspond to 82,744,100 evaluated window
positions. Small/No ELM accounts for roughly three quarters of these
positions, whereas Large ELM represents only about $1.4\%$ and
Type-III ELM is also comparatively sparse. Because stride-one windows
overlap heavily, these counts are not numbers of statistically
independent events; they are reported to characterize the point-wise
imbalance seen by the recognition metric. This imbalance makes the
class-wise Precision--Recall behavior more informative than raw
support-weighted Accuracy.

\noindent\textbf{Type-III ELM.}
The most important minority-class change is observed for Type-III.
Before enhancement, Precision is already high at $84.58\%$, but
Recall is only $52.04\%$, yielding an F1 of $64.44\%$. The temporal
prediction is therefore conservative: Type-III predictions are often
correct when produced, but many true Type-III positions are assigned
to other states. After knowledge enhancement, Recall increases by
$21.40$ points to $73.44\%$, while Precision slightly increases to
$85.26\%$. F1 consequently rises by $14.47$ points to $78.91\%$.
This pattern is consistent with physical and historical context
helping recover Type-III positions that are ambiguous from local
temporal patterns alone, without a corresponding surge in false
Type-III predictions.

\noindent\textbf{L-mode and Small/No ELM.}
L-mode exhibits the largest absolute F1 difference, increasing from
$59.37\%$ to $85.37\%$. Both Precision and Recall improve strongly,
suggesting that explicit confinement-related and physics-aware
information helps distinguish low-confinement intervals from nearby
ELM-related patterns. The dominant Small/No ELM class also improves
from $88.00\%$ to $94.86\%$ F1. Thus, better minority-sensitive
behavior is not obtained simply by sacrificing performance on the
majority class.

\noindent\textbf{Large ELM.}
Large ELM remains the most difficult category. Its Precision improves
from $62.25\%$ to $67.62\%$, whereas Recall decreases slightly from
$68.18\%$ to $67.39\%$, giving a modest F1 increase from $65.08\%$
to $67.50\%$. This result is important because it prevents an
overstated interpretation of minority protection. The protection
mechanism is designed to attenuate harmful knowledge corrections when
the temporal branch already provides a plausible rare-class
hypothesis; it does not force a minority label or guarantee that every
rare class must obtain higher Recall. The remaining Large-ELM gap
also reflects the intrinsic difficulty associated with its extremely
limited support.

Overall, the class-wise behavior complements the aggregate results:
the largest improvements occur where the temporal prediction exhibits
clear confusion, particularly for Type-III and L-mode, while the
dominant class remains strong. At the same time, Large ELM remains a
challenging boundary case, motivating reliability-aware rather than
unrestricted knowledge correction.

\begin{figure*}[!htp]
    \centering
    \includegraphics[width=0.97\textwidth]{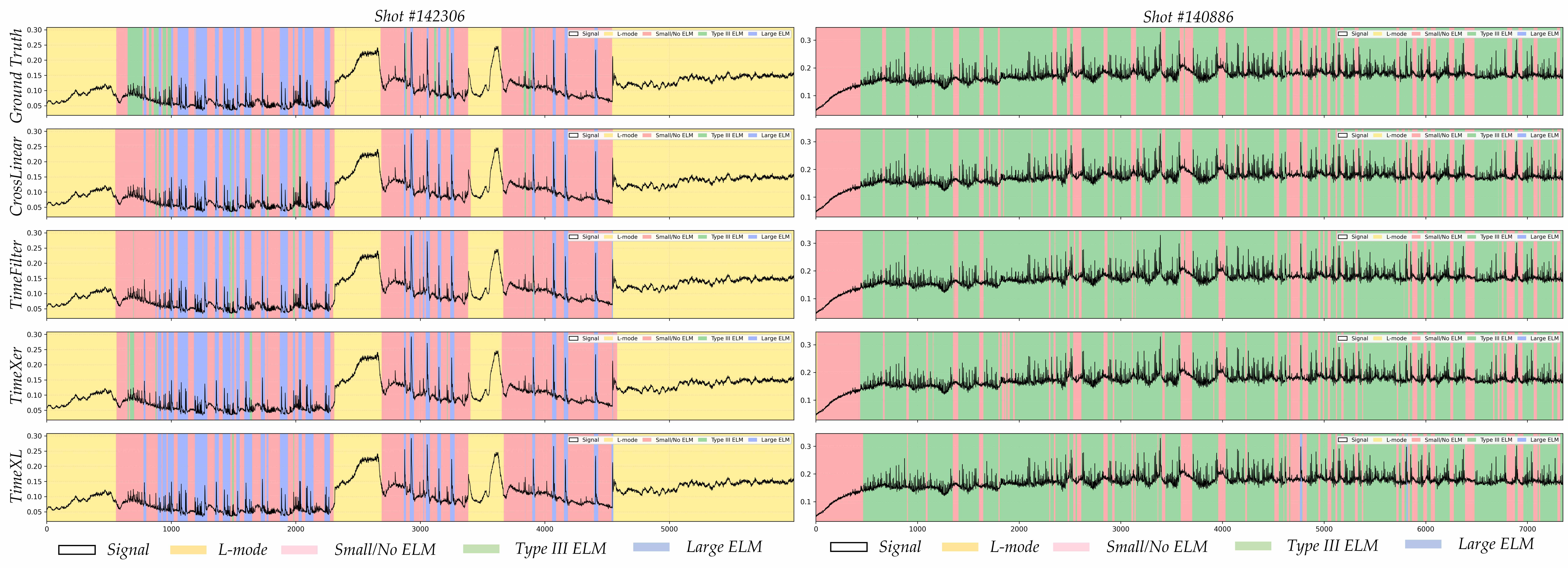}
    \caption{Visualization Result 1. Point-wise ELM recognition results
    of representative temporal backbones equipped with KGCFM.
    Background colors denote the four ELM-related plasma states.}
    \label{fig:visualization_result1}
\end{figure*}

\begin{figure*}[!htp]
    \centering
    \includegraphics[width=0.97\textwidth]{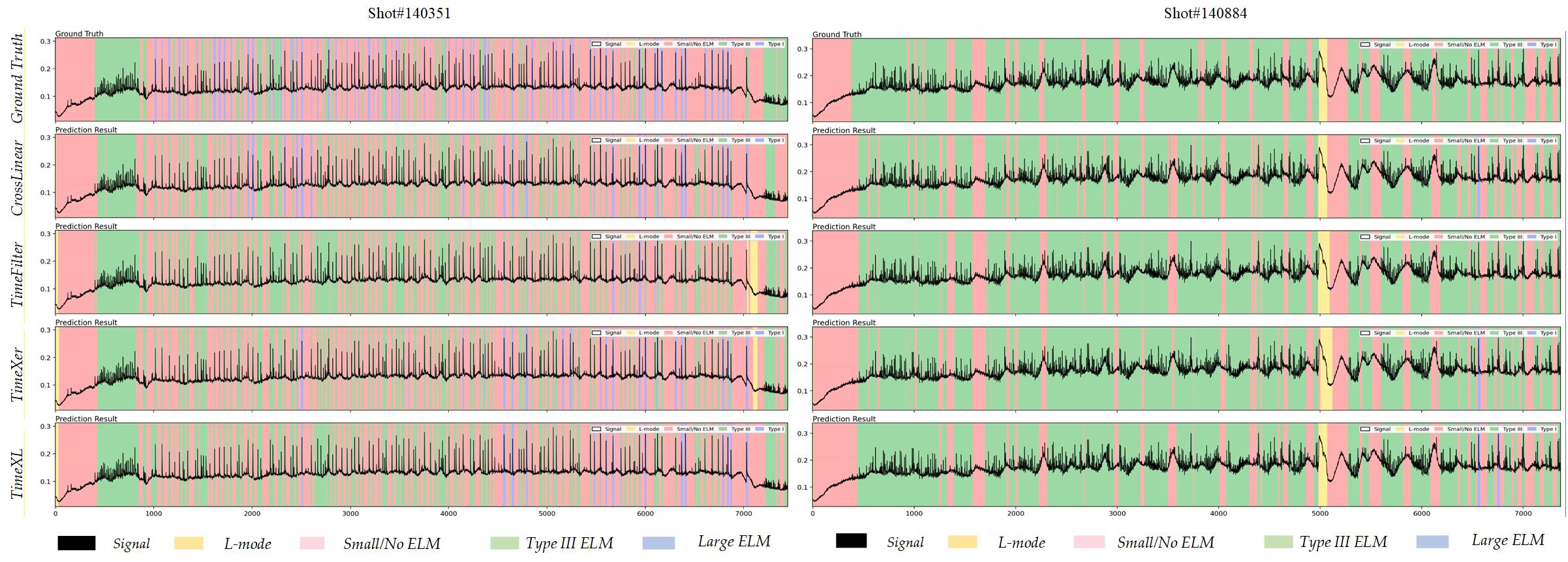}
    \caption{Visualization Result 2. Point-wise ELM recognition results
    of representative temporal backbones equipped with KGCFM.
    Background colors denote the four ELM-related plasma states.}
    \label{fig:visualization_result2}
\end{figure*}

\subsection{Computational Cost}
\label{subsec:computational_cost}

We finally profile the model-side inference cost of the complete
CrossLinear-based KGF-ELM on a single NVIDIA GeForce RTX 4090 with a
batch size of 192. All knowledge pathways remain enabled, including
physical analysis, relational graph encoding, historical retrieval,
feature-level fusion, and decision-level correction. Statistics are
collected over 100 timed batches, corresponding to 19,200 windows.
The profiling results are summarized in
Table~\ref{tab:computational_cost}.

\begin{table}[!t]
\centering
\caption{Batched model-side inference cost of the complete KGF-ELM.}
\label{tab:computational_cost}
\footnotesize
\renewcommand{\arraystretch}{1.08}
\begin{tabular*}{\columnwidth}{@{\extracolsep{\fill}}lr@{}}
\toprule
Measurement & Value \\
\midrule
Profiling windows & 19,200 \\
Batch size & 192 \\
Timed duration & 5.7303\,s \\
Throughput & 3,350.61\,windows/s \\
Average latency & 0.2985\,ms/window \\
Peak allocated GPU memory & 144.89\,MB \\
\bottomrule
\end{tabular*}
\end{table}

Under batched evaluation, the complete model achieves a throughput of
3,350.61 windows/s, corresponding to an average batch-normalized
computation time of 0.2985\,ms/window, with a peak allocated GPU
memory of 144.89\,MB. These results characterize the model-side
computation when the complete knowledge pathway is active and indicate
that large-scale batched evaluation remains practical despite the
additional graph and retrieval operations.

The reported latency should not be interpreted as single-window or
end-to-end EAST deployment latency. It is obtained by dividing the
batched inference time by the number of processed windows and excludes
signal acquisition, synchronization, preprocessing, and communication.
Because a matched temporal-only profiling run is not available under
the same protocol, Table~\ref{tab:computational_cost} reports the
absolute cost of the complete KGF-ELM rather than the incremental
overhead attributable specifically to KGCFM.

\subsection{Visualization}
\label{sec:visualization}

To complement the quantitative evaluation,
Figs.~\ref{fig:visualization_result1} and~\ref{fig:visualization_result2}
present two representative visualizations of point-wise ELM
recognition. The black curves denote the diagnostic signals, while the
colored backgrounds represent the four ELM-related plasma states.
Across different temporal backbones equipped with KGCFM, the dominant
state evolution is consistently recovered, whereas most remaining
disagreements occur around short-duration or rapidly changing regions
where neighboring ELM states exhibit similar temporal patterns.
These visual patterns are consistent with the quantitative results
reported above and show that the knowledge-enhanced models preserve
stable plasma-state regions while most remaining errors are concentrated
around temporally ambiguous transitions.

\subsection{Limitation Analysis}
\label{sec:limitations}

Despite the consistent improvements of KGF-ELM, several limitations remain. First, the current evaluation is restricted to EAST-ELM640, and its generalization to different tokamak devices and operating regimes requires further validation. Second, severe class imbalance remains challenging, particularly for Large ELM, whose limited support restricts the achievable recognition improvement. Third, the present knowledge resource is constructed offline from a fixed subset of training discharges and therefore does not yet adapt dynamically to newly observed plasma conditions. Finally, the reported computational cost measures batched model-side inference and excludes diagnostic acquisition, preprocessing, synchronization, and communication; therefore, it should not be interpreted as a direct demonstration of real-time plasma-control capability.

\section{Conclusion and Future Works}
\label{sec:conclusion}

This paper presented KGF-ELM, a knowledge-guided framework for fine-grained point-wise recognition of four ELM-related plasma states on EAST. By introducing KGCFM, heterogeneous knowledge from physical evidence, ELM prototypes, relational graph representations, and historical discharge cases is collaboratively integrated with data-driven temporal representations through reliability-aware feature- and decision-level fusion. Experiments on EAST-ELM640 demonstrate consistent improvements across
multiple temporal backbones, with the complete model achieving 91.7\%
Accuracy, 81.6\% Macro-F1, and 98.0\% Macro-AUC. Ablation and
sensitivity studies support the complementary roles of different
knowledge sources, while the class-wise and visualization analyses
further characterize the recognition behavior after knowledge
enhancement. These results indicate that structured knowledge can
effectively enhance temporal ELM recognition, particularly for
ambiguous and imbalanced plasma states, without replacing the
underlying data-driven predictor. Future work will focus on extending the framework to broader plasma operating regimes and other tokamak devices, developing dynamically updated knowledge resources, improving recognition of rare ELM states, and exploring lightweight knowledge fusion and end-to-end evaluation toward practical online ELM monitoring.

\FloatBarrier

\bibliographystyle{IEEEtran}
\bibliography{reference}
\end{document}